\documentclass[12pt,preprint]{aastex}

\newcommand{\uv}{\mbox{$u$-$v$}}
\newcommand{\Msol}{\rm{M}_{\sun}}
\newcommand{\Rsol}{\rm{R}_{\sun}}
\newcommand{\Ca}{\mbox{Ca\,\raisebox{+.38ex}{$_{\rm{II}}$}}}

\slugcomment{Accepted to the Astrophysical Journal on 2002 December 12}

\shorttitle{VLBI observations of HR~5110}
\shortauthors{Ransom et al.}

\begin{document}
      
\title{VLBI imaging of the RS~CVn binary star system HR~5110}

\author{R. R. Ransom, N. Bartel and M. F. Bietenholz}
\affil{Department of Physics and Astronomy, York University}
\affil{4700 Keele St., Toronto, Ontario, Canada M3J~1P3} 
\email{rransom@yorku.ca, bartel@yorku.ca,
michael@polaris.phys.yorku.ca}

\author{M. I. Ratner, D. E. Lebach and I. I. Shapiro}
\affil{Harvard-Smithsonian Center for Astrophysics}
\affil{60 Garden St., Cambridge, MA 02138}
\email{mratner@cfa.harvard.edu, dlebach@cfa.harvard.edu,
ishapiro@cfa.harvard.edu}

\author{and J.-F. Lestrade}
\affil{Observatoire de Paris/DEMIRM}
\affil{77 av. Denfert Rochereau, F-75014 Paris, France}
\email{lestrade@mesioa.obspm.fr}

\begin{abstract}

We present VLBI images of the RS~CVn binary star HR~5110 (=BH~CVn;
HD~118216), obtained from observations made at 8.4~GHz on 1994
May~29/30 in support of the NASA/Stanford Gravity Probe B project.
Our images show an emission region with a core-halo morphology.  The
core was $0.39 \pm 0.09$~mas (FWHM) in size, or $66\% \pm 20\%$ of the
$0.6 \pm 0.1$~mas diameter of the chromospherically active K subgiant
star in the binary system.  The halo was $1.95 \pm 0.22$~mas (FWHM) in
size, or $1.8 \pm 0.2$ times the $1.1 \pm 0.1$~mas separation of the
centers of the K and F stars.  (The uncertainties given for the
diameter of the K star and its separation from the F star each reflect
the level of agreement of the two most recent published
determinations.)  The core increased significantly in brightness over
the course of the observations and seems to have been the site of
flare activity that generated an increase in the total flux density of
$\sim$200\% in 12~hours.  The fractional circular polarization
simultaneously decreased from $\sim$10\% to 2.5\%.

\end{abstract}


\section{INTRODUCTION}

HR~5110 (=BH~CVn; HD~118216) is a close binary system with an orbital
period of 2.61~days.  The system consists of an F2~IV star
\citep{sle55} and a K0~IV star \citep{lit86}.  A tabulated list of
selected properties and characteristics of HR~5110 is given in
Table~\ref{thestar}.  HR~5110 is classified as a RS~CVn \citep{hal76},
primarily due to strong \Ca\ H and K emission from the chromosphere of
the cooler K star \citep{con67}, although Little-Marenin et al.\
demonstrate that HR~5110 can also be considered an Algol system seen
very close to pole-on (inclination, $i \lesssim 13\arcdeg$; see
Table~\ref{thestar}).  We adopt the {\it Hipparcos} distance for
HR~5110 of $44.5 \pm 1.3$~pc \citep{esa97} throughout this paper.

Radio emission from HR~5110 is relatively strong and highly variable,
with a flux density that ranges over approximately two orders of
magnitude: Quiescent emission levels are in the 5--20~mJy range
\citep{flo85,wil87}, while flux densities as high as 460~mJy have been
observed during intense flaring episodes \citep{fel83}.  Model fitting
of visibility data from past very-long-baseline interferometry (VLBI)
observations of HR~5110 revealed source sizes ranging from 1.0~mas to
more than 3.7~mas \citep{les84,mut85,lit86}.

Here we report on VLBI observations of HR~5110 made on 1994 May 29/30.
The purpose of the observations was to obtain a position estimate of
HR~5110, which was then a candidate guide star for the NASA/Stanford
relativity gyroscope experiment, Gravity Probe~B \citep[GP-B; see,
e.g.,][]{tur86}.  (HR~5110 was later rejected as the guide star in
favor of a different RS~CVn binary, HR~8703.)  We describe our
observations of HR~5110 in \S\,\ref{obs}, and the data reduction and
analysis procedures in \S\,\ref{reduc}.  In \S\,\ref{results}, we
present our VLBI images of HR~5110 as well as Gaussian model fits to
the visibility data.  In \S\,\ref{discuss}, we discuss our results and
compare them to previous VLBI results for HR~5110.  We summarize our
conclusions in \S\,\ref{concl}.

\placetable{thestar}

\section{OBSERVATIONS \label{obs}}

We made observations of HR~5110 with a global VLBI array of seven
telescopes on 1994~May~29/30 (see Table~\ref{antab}).  Our array had
higher sensitivity and better spatial frequency (\uv) coverage than
previous VLBI observations of this star system.  Our total observing
time was approximately twelve hours.  The observations were centered
at a frequency of 8420~MHz ($\lambda = 3.6$~cm), and were in
right-hand circular polarization (RCP) only, since two of our
antennas, Goldstone and Robledo, did not customarily support
dual-polarization observations and a polarization measurement was not
critical to our support of GP-B.  Data were recorded with Very Long
Baseline Array (VLBA) systems at the Very Large Array (VLA) and the
VLBA-Hancock station and with Mark~III recording systems in the
Mark~IIIA format at all other stations.  The data were correlated at
the Haystack Observatory.

\placetable{antab}

We used the VLA as a stand-alone interferometer (in addition to using
it as an element of our VLBI array) to monitor the variability of the
total flux density and total circular polarization of HR~5110 over the
course of our observations.

We observed HR~5110 in a repeating $\sim$10~minute cycle (including
antenna slew time) with three extragalactic reference sources close on
the sky ($<3\arcdeg$).  Table~\ref{antab} lists the reference sources
observed as well as the dwell time per cycle on each source.  We
scheduled the observations this way so that we could employ phase
referencing and thus determine a position for the radio emission
region of HR~5110 in support of GP-B.  However, once HR~5110 was
dropped as the GP-B guide star, we turned our attention entirely to
the structure and any relative motions of the emission source that we
could discern from a single epoch of observations.  Since the quality
of images generated via hybrid mapping (see \S\,\ref{reduc}) is
superior to that of images generated via phase-reference mapping, we
do not discuss our phase-referencing efforts in this paper.

\section{DATA REDUCTION AND ANALYSIS \label{reduc}}

We analyzed the interferometric data from the VLA using NRAO's
Astronomical Image Processing System (AIPS) and following standard
calibration procedures.  The flux density calibrator source, 3C286,
with flux density defined on the scale of \citet{baa77}, was used to
set the flux density scale for our VLA observations.  We estimate a
standard deviation of 5\% for this absolute flux density calibration.
We used the AIPS program DFTPL to plot the flux density of HR~5110,
time-averaged in 10 minute intervals, over the course of the observing
session.

We also used AIPS to reduce our VLBI data.  The initial reduction,
including fringe-fitting, editing, and preliminary amplitude
calibration, was completed in the usual manner \citep[see,
e.g.,][]{dia95}.  The final calibration of the complex visibilities
and the imaging of HR~5110 was done as follows:

\noindent 1) We first imaged the reference sources using the hybrid
mapping scheme for continuum imaging described in \citet{wal99}.  The
initial model for the self-calibration of each source was a point
source at the phase center, which is the point in the image
corresponding to the position used during the correlation process,
with flux density equal to the total RCP flux density found via the
VLA.  We used the resulting ``final'' images for the three reference
sources to determine the antenna-based amplitude gains as a function
of time over the course of the observations.  For each antenna, the
amplitude gains for the three reference sources agreed to within 5\%
at all times, so we used the average of these three gains to calibrate
the HR~5110 data.

\noindent 2) We then imaged HR~5110 using the same hybrid mapping
scheme except that we performed no amplitude self-calibration since
the star's total flux density varied over the course of the
observations.  The initial model for the self-calibration in phase was
a point source at the phase center.  Note that our procedure discards
information about the absolute position and motion of the source since
the phase self-calibration serves to align the brightest feature with
the phase center.

Flux densities measured with VLBI are not directly tied to the flux
density scale of \citet{baa77} as are those measured with the VLA;
rather, the VLBI flux density scale is based on system temperatures
measured at each antenna over the course of an observing session.  A
comparison of the VLA-determined and VLBI-determined flux densities
for the three extragalactic reference sources that we observed along
with HR~5110 shows that the flux density on the VLBI scale is $15\%
\pm 5\%$ lower than the flux density on the VLA scale, where the
uncertainty has been inferred from the ratios computed for each source
separately.  However, our CLEAN image of HR~5110 based on VLBI data
from the full usable time range (see \S\,\ref{mapsandfits}) contains
$\sim$50\% less flux density than the mean RCP flux density measured
with the VLA during the same hours.  After accounting for the $15\%
\pm 5\%$ difference in flux density scales, we attribute the remaining
difference of $\sim$$35\% \pm 5\%$ between the VLBI CLEAN flux density
and the mean VLA flux density to 1) the nonuniform sampling of the
\uv\ plane with time by our VLBI array over the course of the
observations, and 2) the possible presence of structure that is at
least partially resolved by even the shortest baselines in our VLBI
array.

In addition to producing images of HR~5110, we also fit simple
brightness-distribution models to the VLBI data (averaged into 30
minute bins) for each baseline using the AIPS program OMFIT.  To
describe the structure, we used first a one-component model consisting
of a single elliptical Gaussian.  The center of the elliptical
Gaussian was fixed at the phase center.  Second, we used a
two-component model consisting of a circular Gaussian and an
elliptical Gaussian.  The centers of both the circular and elliptical
Gaussians were fixed at the phase center.  Third, we used a
three-component model consisting of three circular Gaussians.  The
circular Gaussians were initially arranged east-west and the center of
the middle Gaussian was fixed at the phase center.  OMFIT
simultaneously estimates both the fit parameters and the time-varying
antenna-based phase errors (i.e., it self-calibrates in phase).  For
want of any better estimate, the systematic contribution to the
uncertainty in the value of each of the fit parameters for the full
usable data set was estimated by monitoring the values of the
parameter at each iteration of self-calibration.  We chose as the
standard error for each estimated parameter the root-sum-square of the
statistical standard deviation of the fit and the maximum deviation of
that parameter from its final value obtained during the iteration
process.  We also fit the three models described above separately to
each of two subsets of the full usable data set corresponding
approximately to the first half and the second half of the
observations (see \S\,\ref{mapsandfits} for the exact time intervals
for each subset).  To obtain realistic estimates of the total, mostly
systematic, errors for each subset, we repeated the model fit for even
smaller subsets of the data.  Specifically, we divided the data of
each of the above subsets into halves of approximately equal duration
and size (i.e., number of visibilities).  In these cases, we give as
the standard error for each estimated parameter for each subset, the
root-mean-square (rms) of the deviations of the estimates for the two
halves of that subset from the fit value for that same subset.

\section{RESULTS \label{results}}

\subsection{VLA Radio Light Curves \label{lightcurves}}

\begin{figure}
\plotone{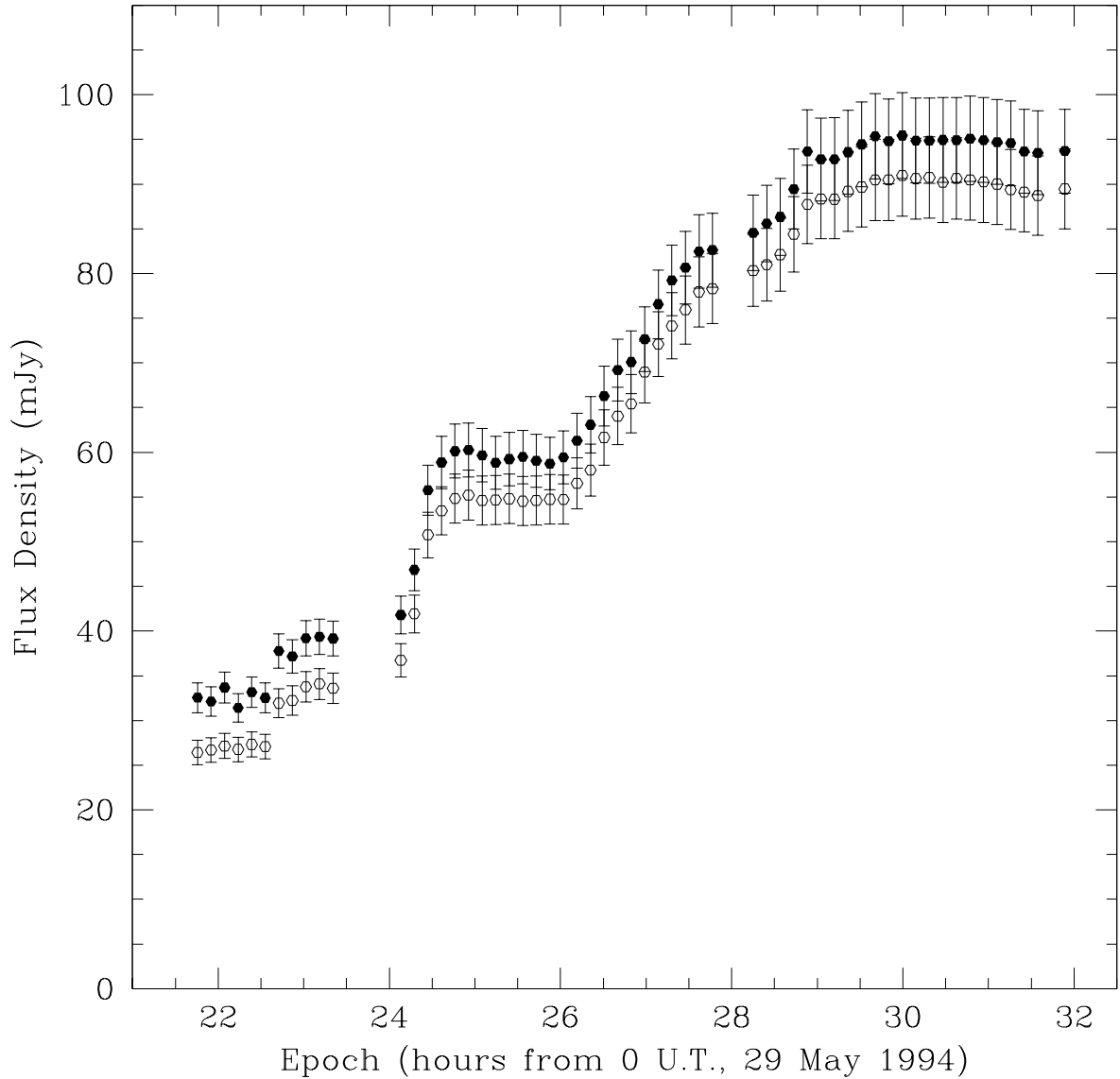}
\figcaption{VLA radio light curve of HR~5110 at 8.4~GHz in RCP (filled
circles) and LCP (open circles) for 1994~May~29/30.  (Note that the
VLBI observations for this session were in RCP only.)  The averaging
interval is 10~minutes, the period of our observing cycle.  The
uncertainties shown include standard errors derived from the internal
scatter of the VLA data but are dominated by the 5\% systematic
contribution that we allowed for the error in flux density calibration
of the VLA data.  The relative accuracy is much higher if, as
expected, the VLA calibration error is not strongly time variable.
\label{hr5110rlc}}
\end{figure}

The VLA radio light curve, presented in Figure~\ref{hr5110rlc}, shows
that HR~5110 had relatively high radio emission during our
1994~May~29/30 observations.  The flux density in RCP increased from
$\sim$33~mJy to $\sim$95~mJy over the course of the observing period.
The light curve also shows that the fractional circular polarization,
defined as $(R-L)/(R+L)$, where $R$ and $L$, respectively, represent
the flux density in RCP and LCP, was $9.6\% \pm 1.4\%$ for the first
hour of observations, and decreased significantly to $2.5\% \pm 0.6\%$
by the last hour.  (As expected, the gain calibration for RCP and LCP
agreed to within 0.2\%, and therefore we believe that the uncertainty
in fractional circular polarization is much smaller than the 5\%
standard error we estimated for the flux density calibration.  The
uncertainties quoted above are thus based solely on the scatter in the
data points during the corresponding hour.)  The sense of circular
polarization, namely right circular, is the same as that observed for
HR~5110 in July 1983 \citep{mut85}.  The constancy of the sense of
circular polarization over time periods of $\sim$10 years has been
observed in two other RS~CVn systems \citep[UX~Arietis and HR~1099;
see, e.g.,][]{mut87}.  The degree of circular polarization in the last
hour of our VLA observations is approximately that observed previously
for HR~5110 in an active emission state ($\sim$2\% with a flux density
of $\sim$140~mJy at 4.9~GHz) by \citet{mut85}.  The observed trend,
namely that the circular polarization decreases with increasing flux
density, is in good agreement with that found for the RS~CVn system
HR~1099 \citep{mut87,mut98,ran02}.

\placefigure{hr5110rlc}

\subsection{VLBI Imaging and Model Fitting \label{mapsandfits}}

\begin{figure}
\plotone{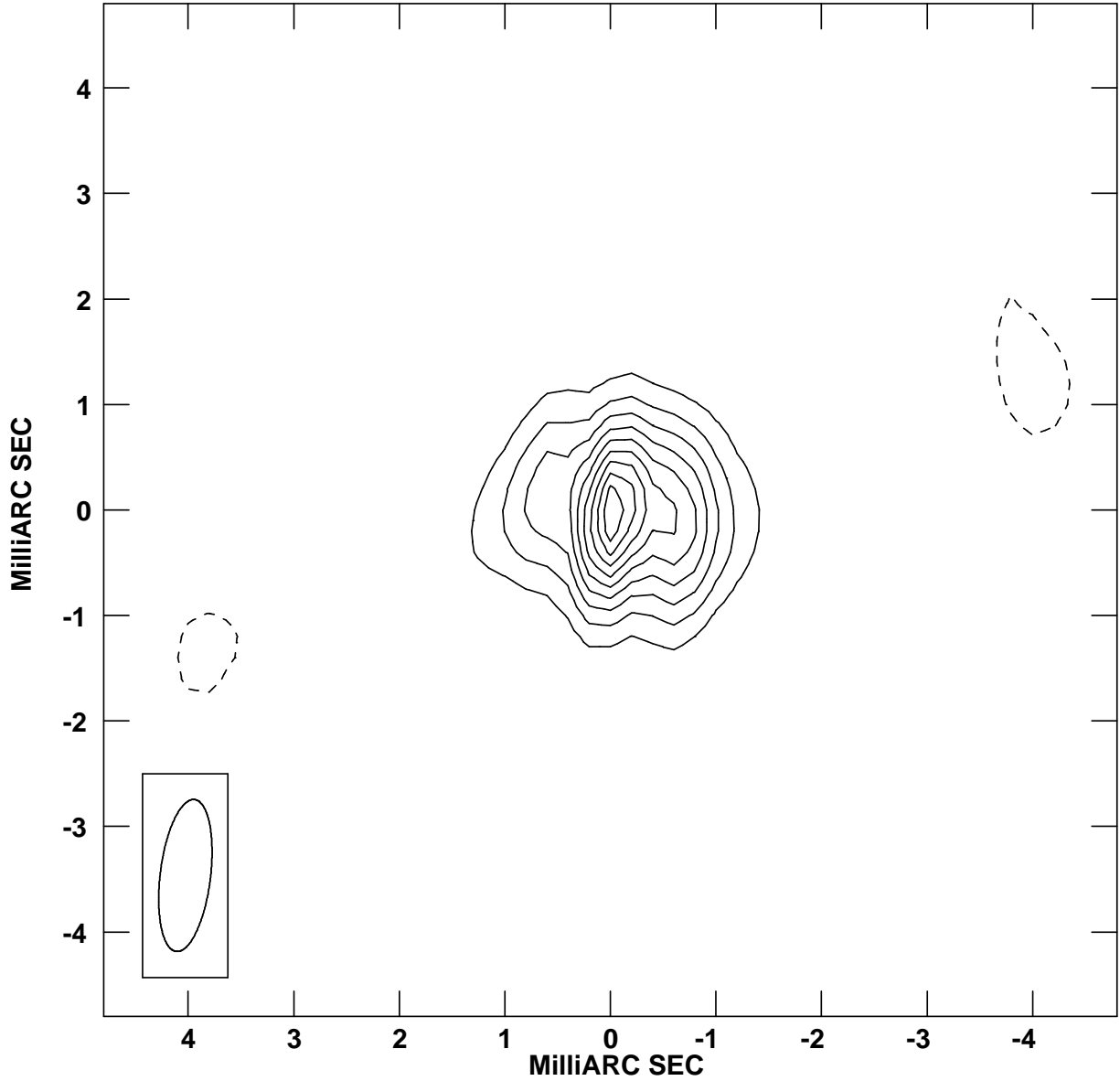}
\figcaption{VLBI image of HR~5110 for 1994 May~29/30.  Here and
hereafter, north is up and east to the left.  The FWHM beam size is
1.5~x~0.5~mas oriented at $-7\arcdeg$ and is shown at the lower left.
The contour levels are -10, 10, 20, 30, 40, 50, 60, 70, 80, and 90\%
of the peak brightness of 14.6~mJy/beam.  The background rms noise
level is 0.47~mJy/beam.  The total CLEAN flux density in the image is
34.1~mJy.
\label{hr5110full}}
\end{figure}

The moderate to high flux density of HR~5110 on 1994~May~29/30 allowed
us to detect the star on all VLBI baselines in most scans ($\sim93\%$)
throughout our observations.  However, for approximately one hour at
both the start and end of our observations, the source was visible at
only two of the seven antennas in our array.  We were not able to
reliably self-calibrate during these times and, therefore, the useful
data set for the imaging process was reduced to a span of ten~hours.
The image made from these ten hours of data (Figure~\ref{hr5110full})
shows that the emission source was extended in the east-west
direction.  Moreover, the concentration of contours around the
brightness peak in the image indicates that there is a compact
emission region within the extended region.  This structure is
consistent with a core-halo morphology.  There is no clear indication
of structure within the compact core.

\placefigure{hr5110full}

The models that we fit to the visibility data yield a quantitative
parameterization of the emission structure seen in the above image.
The one-component model describes the most basic structure of the
source.  Its parameters are listed in Table~\ref{fitsone5110}.  The
$1.66 \pm 0.21$~mas major axis size (full width at half-maximum: FWHM)
and $64\arcdeg \pm 8\arcdeg$ position angle (east of north) of the
elliptical Gaussian is consistent with the footprint of the emission
visible in the image.  However, this model does not describe the
compact region at the phase center.  The two-component model provides
a better fit to the data (i.e., has a 30\% lower reduced chi-square).
The size (FWHM) of the circular Gaussian converged to $0.45 \pm
0.07$~mas, which approximately matches the compact region at the phase
center in the image, while the position angle of the elliptical
Gaussian converged to $70\arcdeg \pm 8\arcdeg$, equal within the
errors to that of the elliptical Gaussian in the one-component model.
However, the value of the axial ratio of the elliptical Gaussian
converged to zero.  Rather than describing a real feature that is
completely unresolved in one dimension, the convergence is more
plausibly related to: 1) the CLEAN beam of the interferometer array
being highly elliptical and oriented approximately perpendicular to
the major axis of the elliptical Gaussian; and 2) the relatively small
number of antennas used for self-calibration.  We therefore do not
consider this model further.  The best fit was obtained with the
three-component model (37\% lower reduced chi-square compared to the
one-component model and 10\% lower reduced chi-square compared to the
two-component model) which is roughly consistent with all the
structure in the image, namely a compact emission region with
additional emission to both east and west.  The parameters for this
model are listed in Table~\ref{fitsthree5110}.  The central component
defines a compact region of size $0.39 \pm 0.09$~mas (FWHM).  The
centers of the east and west components are separated from the center
of the central component by only $\sim$0.6 times their own FWHMs.
They lie along a line in the image plane at a position angle between
approximately 50$\arcdeg$ and 70$\arcdeg$.  Based on the separation of
the most distant parts of the half-maximum contours of its eastern and
western components, we infer that this model defines an extended
emission region $1.95 \pm 0.22$~mas in diameter.

\placetable{fitsone5110}
\placetable{fitsthree5110}

\begin{figure}
\plottwo{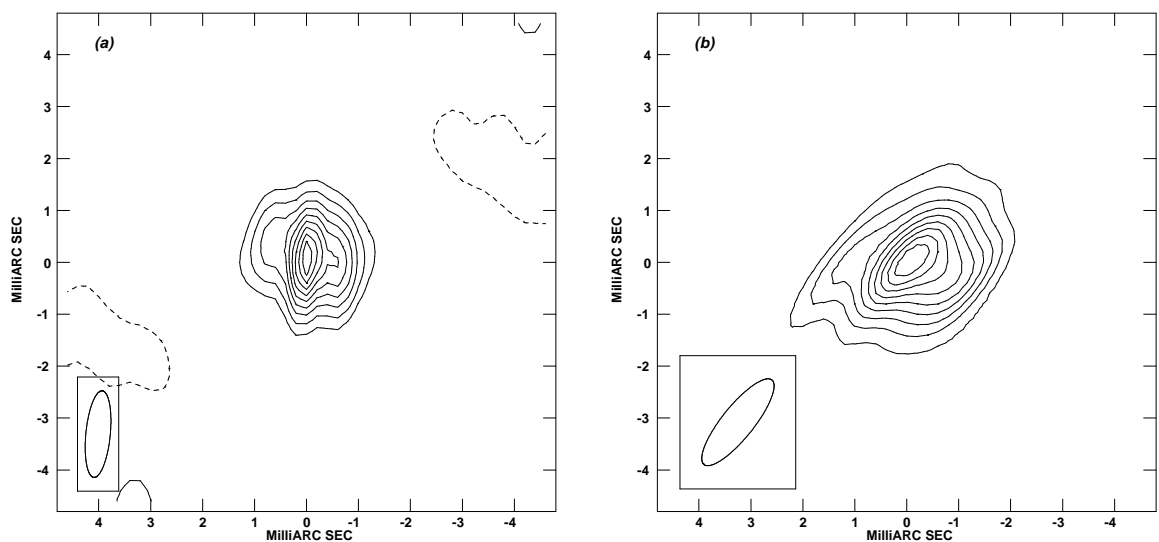}{f3b.eps}
\figcaption{VLBI images of HR~5110 for $(a)$ the first half and $(b)$
the second half of the 1994~May~29/30 observations.  The contour
levels displayed for each image are -10, 10, 20, 30, 40, 50, 60, 70,
80, and 90\% of the beak brightness for that image.  The exact time
interval, FWHM beam size and orientation, peak brightness, background
rms noise, and total CLEAN flux density for each image are:\protect\\
$(a)$ Interval 1: 1994 May~29/22:00--30/02:50~UT; 1.7~x~0.5~mas at
$-6\arcdeg$; 12.6~mJy/beam; $0.50$~mJy/beam; 30.1~mJy. \protect\\
$(b)$ Interval 2: 1994 May~30/02:50--08:00~UT; 2.1~x~0.6~mas at
$-39\arcdeg$; 21.2~mJy/beam; $0.39$~mJy/beam; 57.8~mJy. \protect\\
\label{hr5110snap}}
\end{figure}

To investigate whether there were significant changes in the source
structure of HR~5110 over the course of our observations, we divided
the usable data set, as mentioned in \S\,\ref{reduc}, into two subsets
of nearly equal duration: 1) 1994 May 29, 22:00~UT to May 30,
02:50~UT; and 2) 1994 May 30, 02:50 to 08:00~UT.  We self-calibrated
and imaged the data from each interval independently using the
techniques described in \S\,\ref{reduc}.  The resulting images are
shown in Figure~\ref{hr5110snap}.  Both images are roughly consistent
with the image from the entire data set; i.e., they show a compact
core surrounded by an extended halo.  We also obtained satisfactory
fits of one-component and three-component Gaussian models to the
visibility data for each time interval.  The results for the
one-component models are given in rows~2 and 3 of
Table~\ref{fitsone5110}, and those for the three-component models are
given in rows~2 and 3 of Table~\ref{fitsthree5110}.  For interval 1,
the three-component model fits the data better (i.e., has a 30\% lower
reduced chi-square) than the one-component model.  For interval 2, no
significant improvement (in reduced chi-square) is found with the
three-component model in comparison to the one-component model.  The
three-component models for intervals 1 and 2 are very similar to the
corresponding model for the full time range; i.e., the components lie
along a position angle between approximately 50$\arcdeg$ and
70$\arcdeg$, and define an emission region $\sim$2.0~mas in diameter
(FWHM).  However, while the flux densities of the western and eastern
components remained constant within their standard errors from
interval 1 to interval 2, the flux density of the central component
increased by $13.2 \pm 5.2$~mJy ($92\% \pm 36\%$).  In addition, there
was an increase of $22\arcdeg \pm 14\arcdeg$ from interval 1 to
interval 2 in the position angle of the elliptical Gaussian in the
one-component model.  This change may be indicative of a small
``rotation'' in the source structure.  However, the evidence for
rotation is inconclusive for two reasons: First, no such rotation is
indicated by our three-component models.  Second, our limited
\uv~coverage results in significant changes in the size and
orientation of the beam of the interferometer array from interval 1 to
interval 2.  The fact that the beam computed for the full set of
usable data more closely resembles the beam for interval 1 than that
for interval 2 may explain why our estimates for almost all the model
parameters for the full data set are closer to the corresponding
estimates for interval 1 than to those for interval 2.  If so, the
differences in the beams for the two intervals may also be
contributing significant systematic errors to the apparent changes in
the parameters for the two intervals.  Nevertheless, we doubt that
such systematic errors account for all of the apparent changes.

\placefigure{hr5110snap}

\section{DISCUSSION \label{discuss}}

\begin{figure}
\epsscale{0.99}
\plotone{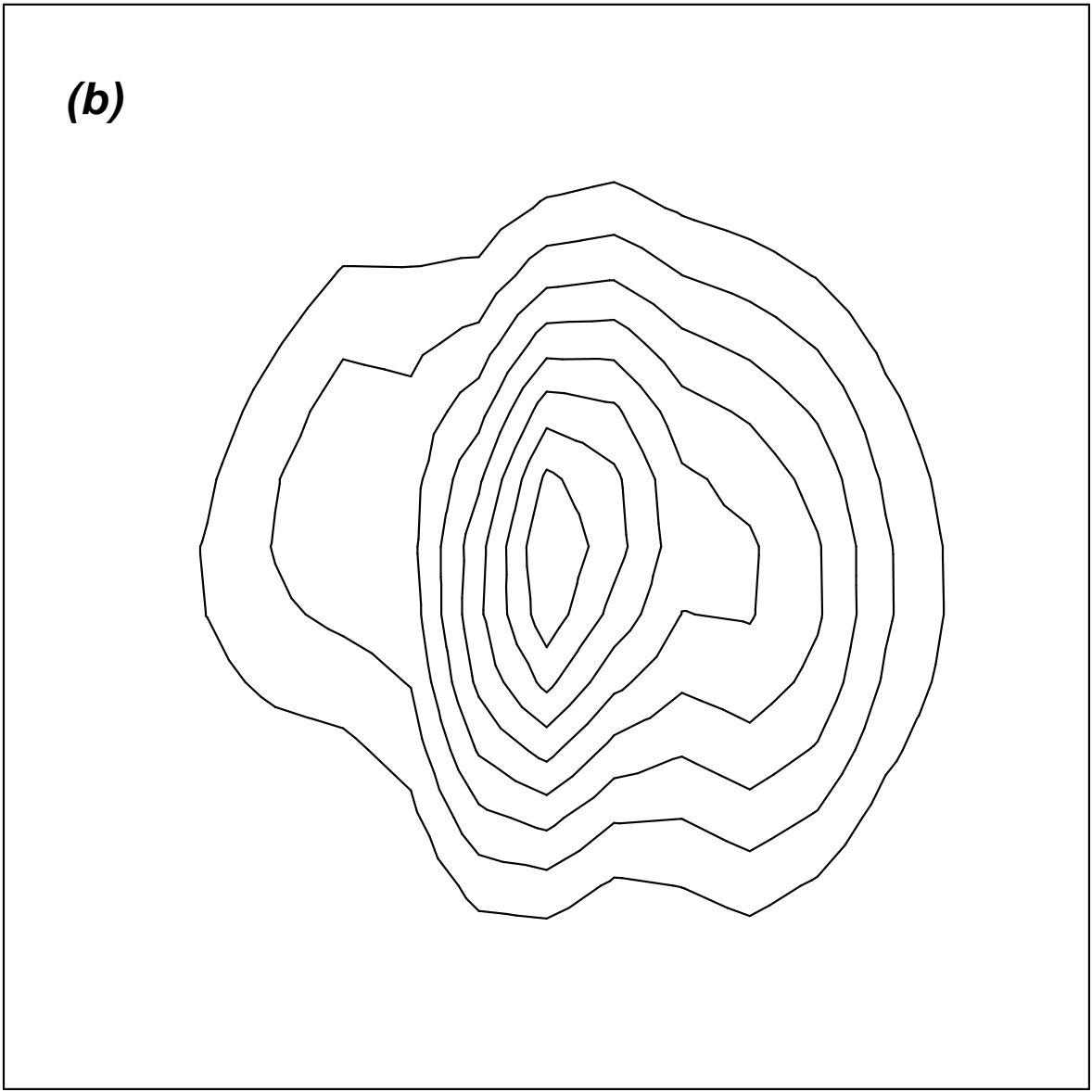}
\figcaption{$(a)$ An artist's three-dimensional rendering of the
HR~5110 binary system as seen from Earth, for the orbital elements and
stellar parameter values in Table~\ref{thestar}.  The scale bar is for
a system distance of 44.5~pc.  The point labeled ``CM'' is the center
of mass of the binary system. The system is shown at the midpoint of
our observations.  From the time of conjunction and the orbital period
given in Table~\ref{thestar}, we calculate the phase to be $0.39 \pm
0.02$.  (Note that conjunction with the F star in front occurs at
phase 0.25.)  The position angle of the line connecting the stars'
centers and starting at the F star is $65\arcdeg$ in this rendering;
i.e., approximately equal to the position angle of the extended
component in our one-component model for the full session.  For this
configuration, the position angle of the ascending node, $\Omega$, is
$-75\arcdeg$ if the stars are rotating counterclockwise and
$-155\arcdeg$ if the stars are rotating clockwise.  The placement of
the K star northeast, rather than southwest, of the F star is
arbitrary, and not required for our inference of $\Omega$.  $(b)$
Figure~\ref{hr5110full} reproduced on the same 3~mas $\times$ 3~mas
scale as $(a)$.
\label{hr5110schem}}
\end{figure}

The images of the HR~5110 emission source from our 1994 May 29/30
observations show a compact region surrounded by an extended,
elongated region.  How do the sizes of these regions compare to the
separation of the stellar components of the HR~5110 binary system and
the diameters of the stars themselves?  The $1.95 \pm 0.22$~mas
angular size we estimated for the extended region from the parameters
of the two outer components in our three-component model (fit to the
full set of usable data) corresponds to a linear size of $(1.3 \pm
0.1) \times 10^{12}$~cm and indicates that the extended region was
$1.8 \pm 0.2$ times larger than the separation\footnote{The binary's
semi-major axis, $a \equiv a_{\rm{F}} + a_{\rm{K}}$, is determined
straightforwardly via the generalization of Kepler's third law: $a =
[P^2 (M_{\rm{F}} + M_{\rm{K}})]^{\frac{1}{3}}$, where $a$ is in A.U.,
$P$ is the period in years, and component masses $M_{\rm{F}}$ and
$M_{\rm{K}}$ are in $\Msol$.  For the purpose of obtaining an angular
separation of the binary components to compare with our model fitting
results, the values of $P$ and $M_{\rm{F}}$ can be taken from
Table~\ref{thestar}, but that of $M_{\rm{K}}$ requires more care,
since inconsistent estimates of the mass ratio $q \equiv
\frac{M_{\rm{K}}}{M_{\rm{F}}}$ have been published by Conti ($0.28 \pm
0.08$, 1967) and Eker \& Doherty ($0.54 \pm 0.01$, 1987).  Rather than
attempting to reconcile the two inconsistent estimates for $q$, we
take as the best estimate of $a$ the average of the $a$ values
computed from each of the two published $q$ values, respectively $9.92
\pm 0.21$~$\Rsol$ and $10.55 \pm 0.03$~$\Rsol$, and take as its
standard error half the difference between these two values.
Converted to angular units by means of the {\it Hipparcos} parallax,
with allowance for the parallax uncertainty, our best estimate of $a$
becomes $1.07 \pm 0.05$~mas.} of the centers of the K and F stars and
$3.3 \pm 0.8$ times larger than the diameter\footnote{Estimates of the
radius of the K star vary between $2.2~\Rsol$ (obtained using the
0.47~mas angular diameter of Little-Marenin et al.\ 1986 and the {\it
Hipparcos} parallax) and $3.4~\Rsol$ \citep{eke87}.  We take as the
best estimate of the diameter of the K star twice the average of these
two radius estimates and take as its standard error the difference
between the two radius estimates.  The resulting best angular diameter
estimate is $0.59 \pm 0.12$~mas.} of the chromospherically active K
star (see Figure~\ref{hr5110schem}{\it a} for a scale diagram of the
HR~5110 binary system).  The $0.39 \pm 0.09$~mas (FWHM) angular size
we estimated for the central component in our three-component model
(fit to the full set of usable data) corresponds to a linear size of
$(2.6 \pm 0.6) \times 10^{11}$~cm, or $66\% \pm 20\%$ of the diameter
of the K star.  The corresponding equivalent brightness
temperature\footnote{We use the formula for the brightness temperature
of a Gaussian source model given in \citet{hje00}.} for the peak of
the central component is $(2.1^{+1.4}_{-0.8}) \times 10^9$~K.  The
corresponding brightness temperatures for the peaks of the east and
west components, respectively, are $(1.9^{+0.6}_{-0.5}) \times 10^8$~K
and $(2.1^{+0.9}_{-0.7}) \times 10^8$~K.  The inferred brightness
temperature for each component is consistent with the gyrosynchrotron
emission mechanism \citep[see, e.g.,][]{dul85}.  Below we discuss the
properties of the radio emission of each of the extended and compact
regions further.

\placefigure{hr5110schem}

How do our results compare with previous VLBI results for HR~5110?
Prior VLBI observations of this system were made at epochs when radio
emission was nearly quiescent \citep{les84}, as well as when the radio
emission was relatively strong \citep{mut85,lit86}.  Lestrade et al.\
observed HR~5110 at 8.4~GHz with a four-element VLBI array on 1982
December~19 when the VLBI-determined flux density of the star was
$\sim$30 mJy.  They found a maximum source size of 1.4~mas (FWHM) by
fitting a single circular Gaussian component to their data.  The
inferred lower bound on the peak brightness temperature of this
component is $2.8 \times 10^8$~K according to the formula of
\citet{hje00}.  Comparing the size and brightness temperature of the
component observed by Lestrade et al.\ to that of the single
elliptical Gaussian component in our one-component model (fit to the
full set of usable data), we find that the $1.66 \pm 0.21$~mas (FWHM)
major axis size of our single component is slightly larger than the
circular size of the single Lestrade et al.\ component, and the
corresponding $(4.3^{+2.2}_{-1.3}) \times 10^8$~K brightness
temperature of our single component is consistent with the lower bound
on the brightness temperature of their single component.  Mutel et
al.\ observed HR~5110 at 5.0~GHz with a six-element array on 1983
July~26 when the total flux density of the star varied between 130 and
170~mJy.  They found a source size of 1.0~mas (FWHM) by fitting a
single circular Gaussian component to their data, which implies a peak
brightness temperature of approximately $7 \times 10^9$~K (again
following Hjellming 2000).  The values for the size and brightness
temperature of this component lie between those we estimated for the
single extended component in our one-component model and the compact
central component in our three-component model, but closer to those
for the central component in our three-component model.
Little-Marenin et al.\ observed HR~5110 at 5.0~GHz with a two-element
VLBI array on 1981 April~6 during a decay phase of a strong radio
flare.  The total flux density during their observation was
approximately 200~mJy.  They estimated a source size of 3.7~mas by
directly comparing the correlated flux density on their single VLBI
baseline to the total flux density from single-dish observations made
at the same time by \citet{fel83} and assuming a single Gaussian
component.  This size is approximately twice as large as that we
measured for our extended region.  Though the size estimate of
Little-Marenin et al.\ is sensitive to any discrepancies between the
flux density scales of the VLBI and single-dish observations, we doubt
that such discrepancies can explain the large difference between the
two measurements.  Thus we believe it probable that there is
significant variability in the size of this region.

How does the source structure we observed for HR~5110 compare to what
has been observed with VLBI for other close binary systems?  Our
images and Gaussian models provide evidence that a radio core-halo
structure was present in HR~5110 during our observations.  The size,
relative to the diameter of the K star, and brightness temperature of
our compact source are each consistent with the values characteristic
of core emission as given by \citet{mut85}.  Similarly, the size,
relative to the binary separation, and brightness temperature of our
extended source are each consistent with the values they give for halo
emission.  A core-halo structure has also been observed for the two
RS~CVn systems UX~Arietis (Mutel et al.) and HR~1099 \citep{ran02} and
for the close binary in Algol (Mutel et al.) during periods when each
system was in a decay phase of a radio flare.  Mutel et al.'s model
for simultaneous core and halo emission describes a system that is in
the decay phase of a flare.  In our case, the core-halo structure was
apparently observed during the onset of a radio flare.  However, we
cannot exclude the possibility that the flare rise we observed is
superimposed upon a decay phase of an earlier flare, because there is
no record of the flux density history of HR~5110 immediately before
our observations.

What can we infer from the time variability of each of the emission
regions identified in our images?  Our three-component models indicate
that, over the $\sim$5~hour period separating the midpoints of
intervals 1 and 2, the flux density of the central component increased
by $13.2 \pm 5.2$~mJy ($92\% \pm 36\%$) while the flux densities of
the eastern and western components remained roughly constant.  In
addition, we note that the sizes of each of the three components
remained essentially unchanged, meaning that the brightness
temperature of only the central component increased, namely about
twofold.  This result suggests that the compact emission region in our
images was located at, or close to, the site of flare activity during
the observations.  However, the full picture is unclear, because the
$17.5 \pm 7.6$~mJy increase in the combined flux density of all three
components in our three-component models does not account for the
$\sim$40~mJy total flux density increase measured with the VLA between
the midpoints of intervals 1 and 2 (even when the VLBI flux density
scale is brought into agreement with the VLA scale).  The discrepancy
is probably due to our model not fully representing the structure of
the source.  Nevertheless, the increase in the flux density of the
compact central component relative to the combined flux density in all
three components of our three-component models (from $\sim$40\% to
$\sim$53\%), together with the observed decrease in the total
fractional circular polarization (from $\sim$10\% to 2.5\%) over the
course of our observations, is consistent with the interpretation that
we are observing a core-halo structure.  The core-halo model used by
\citet{uma93}, to reproduce their observed radio spectra for HR~5110,
predicts that at 8.4~GHz the core emission is optically thick while the
halo emission is optically thin.  Since optically thick
gyrosynchrotron sources are expected to have a low degree of circular
polarization \citep[$\leq$20\% for $\theta \geq 20\arcdeg$ and
$\leq$8\% for $\theta \geq 60\arcdeg$, where $\theta$ is the angle
between the magnetic field and the line of sight to the
observer;][]{dul82}, an increase in the relative strength of the core
emission compared to the halo emission would result in a decrease in
the total fractional circular polarization of the combined source,
consistent with our findings.

Over any $\sim$5~hour period, the two stellar components of the
HR~5110 binary system travel about $30\arcdeg$ along their orbital
paths.  The projection on the sky of the stars' separation vector
rotates by nearly the same value since the system is viewed close to
face-on ($i \lesssim 13\arcdeg$).  The $22\arcdeg \pm 14\arcdeg$
apparent rotation of the extended emission region that we found in our
one-component models (but not in our three-component models) is thus
consistent with the system being tidally locked (see also Ransom et
al.\ 2002 in which the radio structure of HR~1099 was found to
possibly corotate with the stars in that binary system).  If the
observed projected rotation is real, and if the extended emission
region is indeed fixed with respect to the two stars of the binary
system, then we can infer that the direction of rotation of the system
is counterclockwise.  More observations are needed to study a possible
rotation of the radio structure of HR~5110 in detail.

Can we infer the locations with respect to those of the stars within
the HR~5110 binary system of each of the compact and extended emission
regions seen in our images and Gaussian models?  In contrast to
multi-epoch astrometric VLBI studies of close binary systems
\citep[e.g.,][]{les93,les96}, a single VLBI epoch gives no definitive
information about the location within the system of the emission
source.  However, in some cases for a single epoch, the source
structure itself suggests a likely location of emission components
with respect to the stellar components of the binary.  For example, in
the case of the 1996 May~25 observations of HR~1099 by \citet{ran02},
the $\sim$1.7~mas separation and apparent relative rotation of the two
observed compact components suggested two favored alignments within
that binary system, namely emission from well separated regions of the
cooler K star's corona and emission from both the K and G stars.  Our
results for HR~5110 do not favor any particular alignment.  If,
however, we make the reasonable assumption that all the emission we
observed on 1994 May~29/30 is centered on the cooler K star in the
HR~5110 binary system \citep[note that F-type stars have not been
observed to produce radio emission with flux density
$\gtrsim$0.6~mJy;][]{gud95}, then the alignment would be approximately
as illustrated in Figure~\ref{hr5110schem}{\it a,b}.  A possible
scenario which accounts for the compact and extended emission regions
we observed during our observations might then be as follows: Our
observations began near the start of a radio flare that originated
near the surface of the chromospherically active K star.  Particle
acceleration accompanying the flare gave rise to the compact central
component we found in our three-component Gaussian models, namely one
with an angular size of $0.4 \pm 0.1$~mas (FWHM), or $66\% \pm 20\%$
the diameter of the K star.  As our observations continued, additional
particle acceleration by the flare led to an increase in the flux
density of HR~5110 from a nearly quiescent level to levels typical of
active emission periods.  Emission from other energetic particles,
possibly originating from earlier flaring on the K star \citep[see,
e.g.,][]{mut85,fra95}, and now trapped within the extended
magnetosphere of the K star, or, perhaps, within the interconnected
magnetospheres of both the K and F stars, accounts for the two outer
components in our three-component models.

\section{CONCLUSIONS \label{concl}}

\noindent Here we summarize our findings:

\begin{trivlist}

\item{1.} During our observations on 1994 May~29/30, the flux density
of HR~5110 in right circular polarization rose from $\sim$33~mJy to
$\sim$95~mJy.

\item{2.} The radio emission was $9.6\% \pm 1.4\%$ right circularly
polarized during the first hour of observations, and $2.5\% \pm 0.6\%$
right circularly polarized during the last hour.

\item{3.} Our VLBI images suggest an emission region with core-halo
structure, i.e., a compact region approximately at the center of an
extended region.

\item{4.} When we fit a three-component model consisting of three
circular Gaussian components to our data, the estimated diameter of
the central component was $0.39 \pm 0.09$~mas (FWHM), or $66\% \pm
20\%$ of the diameter of the chromospherically active K star in the
binary system.  The diameter of this component remained essentially
unchanged over the course of the observations.  In contrast, its flux
density and brightness temperature nearly doubled during the
observations, suggesting that the central component was at or near the
site of flare activity.

\item{5.} The two outer components in our three-component model define
an extended region $1.95 \pm 0.22$~mas in its longest dimension
oriented at a position angle between $50\arcdeg$ and $70\arcdeg$.
This region is $1.8 \pm 0.2$ times the separation of the centers of
the K and F stars.  The flux densities of each of the outer components
remained roughly constant during the observations.

\end{trivlist}

\acknowledgements

ACKNOWLEDGMENTS.  We thank Adam Jeziak and Jerusha Lederman for help
with the three-dimensional illustration of the HR~5110 binary system.
This research was primarily supported by NASA, through contracts with
Stanford University, and by the Smithsonian Institution and York
University.  Research at York University was partly supported by
NSERC.  NRAO is operated by Associated Universities, Inc., under
cooperative agreement with NSF.  The DSN is operated by JPL/Caltech,
under contract with NASA.  We have made use of the Astrophysics Data
System Abstract Service, created by SAO and supported by NASA.

\clearpage

\begin{deluxetable}{l c c@{}c c c}
\tabletypesize{\scriptsize}
\tablecaption{Properties and Orbital Elements of HR~5110\label{thestar}}
\tablewidth{0pt}
\tablehead{
  \colhead{Parameter} &
  \colhead{} &
  \multicolumn{2}{c}{Value} &
  \colhead{} &
  \colhead{Reference}
}
\startdata
RA\tablenotemark{a}\, (h m s; J2000)                         &\phm{space} & \multicolumn{2}{c}{13 34 47.74631$\pm$0.000036}       &\phm{space} & 1 \\
Dec\tablenotemark{a}\, ($\arcdeg$ $\arcmin$ $\arcsec$; J2000)&\phm{space} & \multicolumn{2}{c}{37 10 56.77936$\pm$0.00040}        &\phm{space} & 1 \\
Trig. Parallax (mas)                                         &\phm{space} & \multicolumn{2}{c}{22.46$\pm$0.62,\ 22.21$\pm$0.45}   &\phm{space} & 1,2 \\
Distance (pc)                                                &\phm{space} & \multicolumn{2}{c}{44.5$\pm$1.3,\ 45.0$\pm$0.9}       &\phm{space} & 1,2 \\
\noalign{\vspace{10pt}}
Orbital Elements:\tablenotemark{b}                           &\phm{space} &                           &                           &\phm{space} & \\
$a \sin i$ ($10^{11}$ cm) \tablenotemark{c}                  &\phm{space} & \phm{++}0.40\phm{--}      & \phm{++}0.75\phm{--}      &\phm{space} & 3 \\
$a \sin i$ (mas) \tablenotemark{d}                           &\phm{space} & \phm{++}0.06\phm{--}      & \phm{++}0.11\phm{--}      &\phm{space} & \\
$P$ (days)                                                   &\phm{space} & \multicolumn{2}{c}{2.613214$\pm$0.000003}             &\phm{space} & 4 \\
$i$ ($\arcdeg$)                                              &\phm{space} & \multicolumn{2}{c}{8.9}                               &\phm{space} & 3 \\
$e$                                                          &\phm{space} & \multicolumn{2}{c}{0}                                 &\phm{space} & 3 \\
$\omega$                                                     &\phm{space} & \multicolumn{2}{c}{--}                                &\phm{space} & \\
$\Omega$                                                     &\phm{space} & \multicolumn{2}{c}{(Not Determined)}                  &\phm{space} & \\
$T_{conj}$ (HJD) \tablenotemark{e}                           &\phm{space} & \multicolumn{2}{c}{2445766.655$\pm$0.013}             &\phm{space} & 3 \\
\noalign{\vspace{10pt}}
Stellar Properties:\tablenotemark{b}                         &\phm{space} &                           &                           &\phm{space} & \\
Spectral Type                                                &\phm{space} & \phn F2~IV                & \phn K0~IV                &\phm{space} & 5,6\tablenotemark{f} \\
Mass ($\Msol$)                                               &\phm{space} & \phn 1.5                  & \phn 0.8\tablenotemark{c} &\phm{space} & 3 \\
Radius ($\Rsol$)                                             &\phm{space} & \phn 2.6\tablenotemark{g} & \phn 3.4\tablenotemark{h} &\phm{space} & 3 \\
Radius (mas) \tablenotemark{d}                               &\phm{space} & \phn 0.27                 & \phn 0.36                 &\phm{space} & \\
\enddata
\tablerefs{
  1. {\it Hipparcos} Catalogue \citep{esa97};\phn
  2. VLBI \citep{les99};\phn
  3. Eker \& Doherty 1987;\phn
  4. Mayor \& Mazeh 1987;\phn
  5. Slettebak 1955;\phn
  6. Little-Marenin et al.\ 1986
  }
\tablenotetext{a}{The {\it Hipparcos} (J2000) coordinates are for
epoch 1991.25.}
\tablenotetext{b}{Two entries correspond to the two stars of the
binary system.}
\tablenotetext{c}{Using $1.5~\Msol$ for the mass of the F star, and
the mass ratio $q \equiv \frac{M_{\rm{K}}}{M_{\rm{F}}} = 0.54 \pm
0.01$ \citep{eke87}, we then obtain the mass of the K star. We note
here that Conti 1967 found a significantly different value for the
mass ratio of the system: $q = 0.28 \pm 0.08$.  This value leads to a
larger estimate of the inclination of the orbit, $i = 13.2\arcdeg$,
but does not significantly affect the semi-major axis of the binary,
$a = a_{F} + a_{K}$, since $a \propto (M_{\rm{F}} +
M_{\rm{K}})^{\frac{1}{3}}$.}
\tablenotetext{d}{For a distance of 44.5~pc.}
\tablenotetext{e}{Heliocentric time of conjunction with the K star in
front.}
\tablenotetext{f}{Note also inference of K2~IV by Eker \& Doherty
1987.}
\tablenotetext{g}{Modified using the {\it Hipparcos} parallax.}
\tablenotetext{h}{Radius of the K star in the limit of filling its
Roche lobe.  We note that the mass ratio, $q = 0.28 \pm 0.08$, found
by Conti 1967 leads to a smaller value for the Roche lobe radius of
the K star: $2.6~\Rsol$.  This radius is more consistent with the
angular diameter of the K star found by Little-Marenin et al.\ 1986:
0.47~mas.}
\end{deluxetable}

\begin{deluxetable}{r@{ }l@{ }l@{\protect\phantom{xxx}} c c c r}
\tabletypesize{\footnotesize}
\tablecaption{VLBI Observations of HR~5110\label{antab}}
\tablewidth{0pt}
\tablehead{
  \multicolumn{3}{c}{Observing} &
  \colhead{Total Time\tablenotemark{a}} &
  \colhead{Source} &
  \colhead{Dwell Time} &
  \colhead{Antennas\tablenotemark{b}}\\
  \multicolumn{3}{c}{Date} &
  \colhead{(hours)} &
  \colhead{} &
  \colhead{(min. per 10 min. cycle)} &
  \colhead{}
}
\startdata
 1994&May&29/30 & 12.8 & HR~5110   & 2.0  & Aq,Eb,Gb,Go,Ro,Y27,Hn \\
     &   &      &      & J1340+379 & 1.3  & \\
     &   &      &      & J1334+371 & 1.7  & \\
     &   &      &      & J1328+363 & 0.8  & \\
\enddata
\tablenotetext{a}{Total length of the observing run.}
\tablenotetext{b}{
  Aq =  46m, ISTS (now CRESTech/York Univ.), Algonquin Park, Ontario, Canada;\phn
  Eb = 100m, MPIfR, Effelsberg, Germany;\phn
  Gb =  43m, NRAO, Green Bank, WV, USA;\phn
  Go =  70m, NASA-JPL, Goldstone, CA, USA;\phn
  Ro =  70m, NASA-JPL, Robledo, Spain;\phn
  Y27 = phased VLA equivalent diameter 130m, NRAO, near Socorro, NM, USA;\phn
  Hn =  25m, NRAO, Hancock, NH, USA (one of the 10 antennas comprising the VLBA).
  }
\end{deluxetable}

\begin{deluxetable}{c c r c r c r c r}
\tabletypesize{\scriptsize}
\tablecaption{One-Component Model Parameters for HR~5110\label{fitsone5110}}
\tablewidth{0pt}
\tablehead{
  \colhead{} &
  \colhead{} &
  \multicolumn{7}{c}{Elliptical Gaussian\tablenotemark{a}} \\
  \colhead{Interval} &
  \colhead{} &
  \colhead{$S$} &
  \colhead{} &
  \colhead{Major Axis} &
  \colhead{} &
  \colhead{Axial Ratio} &
  \colhead{} &
  \colhead{p.a.} \\
  \colhead{(UT)} &
  \colhead{} &
  \colhead{(mJy)} &
  \colhead{} &
  \colhead{(mas)} &
  \colhead{} &
  \colhead{} &
  \colhead{} &
  \colhead{(deg)} \\
  \hline
  \colhead{(1)} &
  \colhead{} &
  \colhead{(2)} &
  \colhead{} &
  \colhead{(3)} &
  \colhead{} &
  \colhead{(4)} &
  \colhead{} &
  \colhead{(5)}
}
\startdata
full         &\phm{s} &$39.6 \pm 1.9$ &\phm{s} &$1.66 \pm 0.21$ &\phm{s} &$0.58 \pm 0.17$ &\phm{s} &$64 \pm \phn8$ \\
22:00--26:50 &\phm{s} &$33.0 \pm 2.5$ &\phm{s} &$1.67 \pm 0.24$ &\phm{s} &$0.59 \pm 0.21$ &\phm{s} &$63 \pm 10$ \\
26:50--32:00 &\phm{s} &$56.4 \pm 4.9$ &\phm{s} &$1.77 \pm 0.23$ &\phm{s} &$0.58 \pm 0.21$ &\phm{s} &$85 \pm 10$ \\
\enddata
\tablenotetext{a}{The center of the elliptical Gaussian component is
fixed at the phase center.}
\tablecomments{The description of the column entries is given below.
The standard errors were estimated using the procedure described in
\S\,\ref{reduc}.\\
(1) Data interval with respect to 0 UT, 1994 May~29 (full $\equiv$
    full usable time range, 22:00--32:00~UT).\\
(2) Flux density of the elliptical Gaussian component.\\
(3) FWHM of the major axis of the Gaussian component.\\
(4) Axial ratio of the Gaussian component.\\
(5) Position angle (east of north) of the major axis of the Gaussian component.\\
}
\end{deluxetable}

\begin{deluxetable}{c r r r r r r r r r r}
\tabletypesize{\scriptsize}
\rotate
\tablecaption{Three-Component Model Parameters for HR~5110\label{fitsthree5110}}
\tablewidth{0pt}
\tablehead{
  \colhead{} &
  \multicolumn{2}{c}{Center\tablenotemark{a}} &
  \multicolumn{4}{c}{West} &
  \multicolumn{4}{c}{East} \\
  Interval &
  \multicolumn{1}{c}{$S$} &
  \multicolumn{1}{c}{FWHM} & 
  \multicolumn{1}{c}{$S$} &
  \multicolumn{1}{c}{FWHM} &
  \multicolumn{1}{c}{Rad} &
  \multicolumn{1}{c}{$\theta$} &
  \multicolumn{1}{c}{$S$} &
  \multicolumn{1}{c}{FWHM} &
  \multicolumn{1}{c}{Rad} &
  \multicolumn{1}{c}{$\theta$} \\
  (UT) &
  \multicolumn{1}{c}{(mJy)} &
  \multicolumn{1}{c}{(mas)} &
  \multicolumn{1}{c}{(mJy)} &
  \multicolumn{1}{c}{(mas)} &
  \multicolumn{1}{c}{(mas)} &
  \multicolumn{1}{c}{(deg)} &
  \multicolumn{1}{c}{(mJy)} &
  \multicolumn{1}{c}{(mas)} &
  \multicolumn{1}{c}{(mas)} &
  \multicolumn{1}{c}{(deg)} \\
  \hline
  (1) &
  \multicolumn{1}{c}{(2)} &
  \multicolumn{1}{c}{(3)} &
  \multicolumn{1}{c}{(4)} &
  \multicolumn{1}{c}{(5)} &
  \multicolumn{1}{c}{(6)} &
  \multicolumn{1}{c}{(7)} &
  \multicolumn{1}{c}{(8)} &
  \multicolumn{1}{c}{(9)} &
  \multicolumn{1}{c}{(10)} &
  \multicolumn{1}{c}{(11)}
}
\startdata
full         &$18.6 \pm 2.7$ &$0.39 \pm 0.09$ &$10.3 \pm 2.2$ &$0.91 \pm 0.13$ &$0.51 \pm 0.10$ &$-118 \pm \phn7$ &$\phn9.5 \pm 1.3$ &$0.92 \pm 0.11$ &$0.50 \pm 0.10$ &$58 \pm \phn9$ \\
22:00--26:50 &$14.4 \pm 2.8$ &$0.39 \pm 0.09$ &$10.8 \pm 2.5$ &$0.93 \pm 0.13$ &$0.49 \pm 0.13$ &$-119 \pm \phn9$ &$\phn9.6 \pm 2.6$ &$0.92 \pm 0.16$ &$0.53 \pm 0.13$ &$59 \pm 11$ \\
26:50--32:00 &$27.6 \pm 4.4$ &$0.40 \pm 0.10$ &$14.4 \pm 3.4$ &$0.88 \pm 0.14$ &$0.54 \pm 0.14$ &$-122 \pm 12$    &$10.3 \pm 2.5$    &$0.87 \pm 0.20$ &$0.46 \pm 0.14$ &$55 \pm 11$ \\
\enddata
\tablenotetext{a}{The center of the central circular Gaussian
component is fixed at the phase center.}
\tablecomments{The description of the column entries is given below.
The standard errors were estimated using the procedure described in
\S\,\ref{reduc}. \\
(1) Data interval with respect to 0 UT, 1994 May~29 (full $\equiv$
    full usable time range, 22:00--32:00~UT).\\
(2) Flux density of the central circular Gaussian component. \\
(3) Size (FWHM) of the central circular Gaussian component. \\
(4) Flux density of the western circular Gaussian component. \\
(5) Size (FWHM) of the western circular Gaussian component. \\
(6,7) Location (radial distance and position angle, respectively) of
the western circular Gaussian component. \\
(8) Flux density of the eastern circular Gaussian component. \\
(9) Size (FWHM) of the eastern circular Gaussian component. \\
(10,11) Location (radial distance and position angle, respectively) of
the eastern circular Gaussian component. \\
}
\end{deluxetable}

\clearpage

\begin{figure}
\plotone{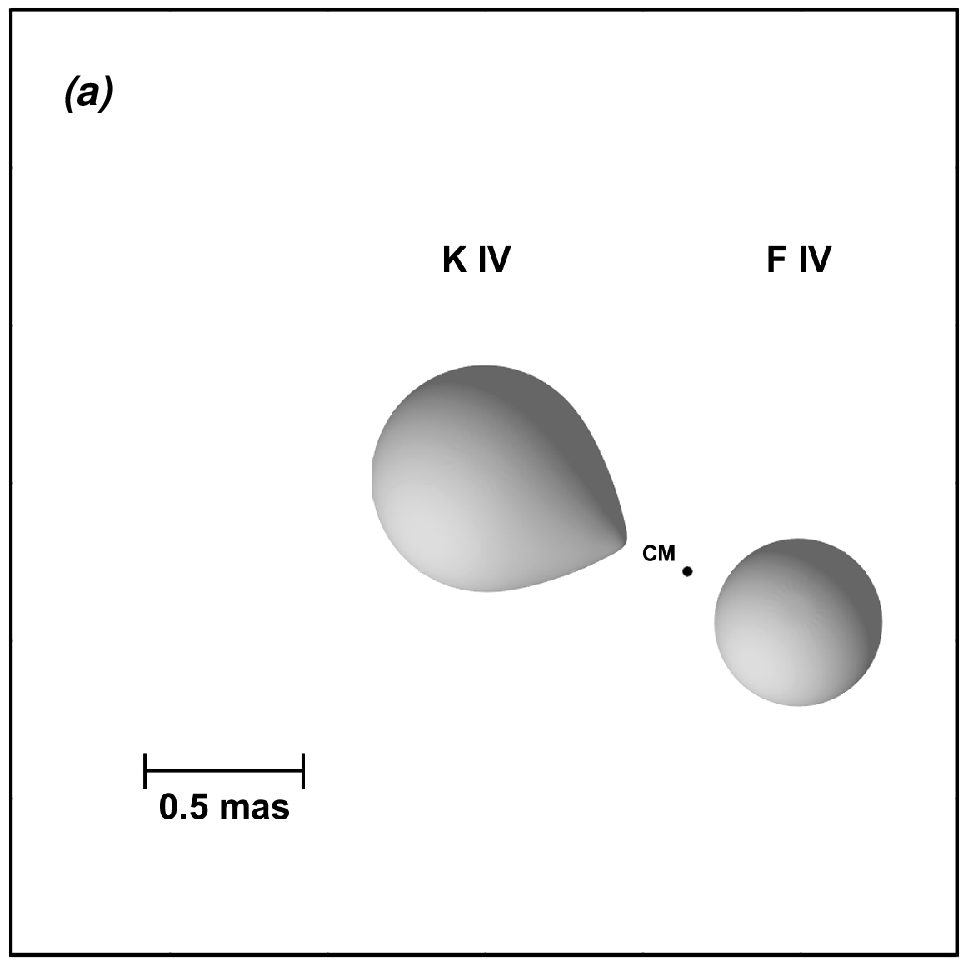}
\end{figure}


\clearpage

\begin{thebibliography}{}

\bibitem[Baars et al.(1977)]{baa77} Baars, J. W. M., Genzel, R.,
    Pauliny-Toth, I. I. K., \& Witzel, A.  1988, \aap, 61, 99
\bibitem[Conti(1967)]{con67} Conti, P. S.  1967, \apj, 149, 629
\bibitem[Diamond(1995)]{dia95} Diamond, P. J.  1995, in ASP Conf. Ser. 82, 
    Very Long Baseline Interferometry and the VLBA, ed. J. A. Zensus,
    P. J. Diamond, \& P. J. Napier (San Francisco: ASP), 227
\bibitem[Dulk(1985)]{dul85} Dulk, G. A.  1985, \araa, 23, 169
\bibitem[Dulk \& Marsh(1982)]{dul82} Dulk, G. A., \& Marsh, K. A.
    1982, \apj, 259, 350
\bibitem[Eker \& Doherty(1987)]{eke87} Eker, Z., \& Doherty, L. R.
    1987, \mnras, 228, 869
\bibitem[Feldman(1983)]{fel83} Feldman, P. A.  1983, in IAU
    Colloq. 71, Activity in Red-Dwarf Stars, ed. P. B. Byrne \& M. Rodono
    (Dordrecht: Reidel), 407
\bibitem[Florkowski et al.(1985)]{flo85} Florkowski, D. R., Johnston,
    K. J., Wade, C. M., \& de Vegt, C.  1985, \aj, 90, 2381
\bibitem[Franciosini \& Chiuderi-Drago(1995)]{fra95} Franciosini, E.,
    \& Chiuderi-Drago, F.  1995, \aap, 297, 535
\bibitem[G\"{u}del, Schmitt, \& Benz(1995)]{gud95} G\"{u}del, M.,
    Schmitt, J. H. M. M., \& Benz, A. O.  1995, \aap, 302, 775
\bibitem[Hall(1976)]{hal76} Hall, D. S.  1976, in IAU Colloquium 29, 
    Multiple Periodic Variable Stars, ed. W. S. Fitch (Dordrecht: Reidel), 287
\bibitem[Hjellming(2000)]{hje00} Hjellming, R. M.  2000, in Allen's
    Astrophysical Quantities, ed. A. N. Cox (4th ed.; New York: Springer),
    121
\bibitem[Lestrade et al.(1984)]{les84} Lestrade, J.-F., Mutel, R. L.,
    Preston, R. A., Scheid, J. A., \& Phillips, R. B.  1984, \apj, 279, 184
\bibitem[Lestrade et al.(1993)]{les93} Lestrade, J.-F., Phillips, R. B.,
    Hodges, M. H., \& Preston, R. A.  1993, \apj, 410, 808
\bibitem[Lestrade(1996)]{les96} Lestrade, J.-F.  1996, in IAU Symposium no. 176, 
    Stellar Surface Structure, ed. K. G. Strassmeier \& J. L. Linsky 
    (Dordrecht: Kluwer), 173
\bibitem[Lestrade et al.(1999)]{les99} Lestrade, J.-F., et al.  1999, \aap,
    344, 1014
\bibitem[Little-Marenin et al.(1986)]{lit86} Little-Marenin, I. R., Simon, T.,
    Ayres, T. R., Cohen, N. L., Feldman, P. A., Linsky, J. L., Little, S. J.,
    \& Lyons, R.  1986, \apj, 303, 780
\bibitem[Mayor \& Mazeh(1987)]{may87} Mayor, M., \& Mazeh, T.  1987,
    \aap, 171, 157
\bibitem[Mutel et al.(1985)]{mut85} Mutel, R. L., Lestrade, J.-F.,
    Preston, R. A., \& Phillips, R. B.  1985, \apj, 289, 262
\bibitem[Mutel et al.(1987)]{mut87} Mutel, R. L., Morris, D. H., Doiron, D. J.,
    \& Lestrade, J.-F.  1987, \aj, 93, 1220
\bibitem[Mutel et al.(1998)]{mut98} Mutel, R. L., Molnar, L. A., Waltman, E. B.,
    \& Ghigo, F. D.  1998, \apj, 507, 371
\bibitem[Perryman(1997)]{esa97} Perryman, M. A. C.  1997, The {\it Hipparcos}
    and Tycho Catalogues (Noordwijk: ESA)
\bibitem[Ransom et al.(2002)]{ran02} Ransom, R. R., Bartel, N.,
    Bietenholz, M. F., Ratner, M. I., Lebach, D. E., Shapiro, I. I., \&
    Lestrade, J. F.  2002, \apj, 572, 487
\bibitem[Slettebak(1955)]{sle55} Slettebak, A.  1955, \apj, 121, 653
\bibitem[Turneaure, Everitt, and Parkinson(1986)]{tur86} Turneaure, J. P.,
    Everitt, C. W. F., \& Parkinson, B. W.  1986, in Proc. Fourth Marcel
    Grossman Meeting on General Relativity, ed. R. Ruffini 
    (Amsterdam: Elsevier), 411
\bibitem[Umana et al.(1993)]{uma93} Umana, G., Trigilio, C.,
    Hjellming, R. M., Catalano, S., \& Rodon\'{o}, M.  1999, \aap, 267, 126
\bibitem[Walker(1999)]{wal99} Walker, R. C.  1999, in ASP Conf. Ser. 180,
   Synthesis Imaging in Radio Astronomy II, ed. G. B. Taylor, C. L. Carilli,
   \& R. A. Perley (San Francisco: ASP), 433
\bibitem[Willson \& Lang(1987)]{wil87} Willson, R. F., \& Lang, K. R.
   1987, \apj, 312, 278

\end{thebibliography}
\end{document}